\documentclass[fleqn,12pt,twoside]{article}
\usepackage[headings]{espcrc1}

\readRCS
$Id: espcrc1.tex,v 1.2 2004/02/24 11:22:11 spepping Exp $
\usepackage{graphicx}
\usepackage[figuresright]{rotating}

\newcommand{\AmS}{{\protect\the\textfont2
  A\kern-.1667em\lower.5ex\hbox{M}\kern-.125emS}}

\title{Comment on ``Precise half-life values for two-neutrino double beta decay"}

\author{B.\ Pritychenko\address[BNLab]{National Nuclear Data Center, Brookhaven National Laboratory, Upton, NY 11973-5000, U.S.A.}
}
       
\runtitle{Comment on ``Precise half-life values for two-neutrino double beta decay"}
\runauthor{B. Pritychenko}

\begin{document}

\maketitle

\begin{abstract}
The results by A.S. Barabash [Phys. Rev. {\bf C 81}, 035501 (2010)] are shown to be incomplete. 
Tellurium data sets were reanalyzed using the best practices of ENSDF evaluations. 
Present analysis indicates much higher value for the 2$\nu$-mode half-life time in $^{128}$Te and 
the corresponding ratio of $^{128,130}$Te half-lives. These values imply the \mbox{$T^{2 \nu}_{1/2}$ $\sim$ $\frac{1}{E^{8}}$}  
$\beta\beta$-decay transition energy trend that is consistent with the two-nucleon mechanism of  $\beta\beta$(2$\nu$)-decay.
\end{abstract}

\section{$\beta\beta$-decay Data Analysis}
For 75 years double beta decay fascinates the hearts and minds of many scientists \cite{35Go} due to the possible implications on 
nuclear physics and fundamental symmetries. During these years experimental $\beta\beta$-decay research was based on 
two complimentary approaches: direct and geochemical measurements. In direct experiments, 2$\nu$- and 0$\nu$-mode decay data 
are accumulated online over few years and analyzed, while in the geochemical experiments, scientists go through extensive 
chemical analysis of rock specimen to extract daughter nuclei due to 2$\nu$+0$\nu$ modes of $\beta\beta$-decay.

In the recent work \cite{10Ba} many of important experimental results are extensively compiled and recommended T$_{1/2}$ 
are deduced using Particle Data Group procedures \cite{PDG}. One of the most interesting cases is related to the analysis 
of $^{128,130}$Te data sets. In the analysis, author goes through extensive selection, removal of discrepant data sets 
and adjustments procedures for $^{128}$Te geochemical data sets using his previous work \cite{00Ba} on time variation of 
weak interaction constant as an explanation. 

In fact, geochemical experiments are very difficult to perform because many parameters such as exact age of the specimen, 
its geological history, etc. are out of experimentalist control and disagreement between two groups of geochemical 
results is not unusual. The discrepancies can be found even among direct measurements of $^{76}$Ge $\beta\beta$(2$\nu$) half-lives performed by 
ITEP/Yerevan,  Heidelberg-Moscow and IGEX collaborations \cite{90Va,90Mi,01Kl,05Ba,02Aa}. Additionally, one cannot reject $^{128}$Te T$_{1/2}$ value 
from the Washington University group \cite{93Be} but still use the $^{128}$Te/$^{130}$Te ratio from the same group. 
Such selective rejection of the legitimate data sets is in direct contradiction with the best practices of nuclear structure evaluations \cite{ENSDF}.

The recommended $\beta\beta$-decay half-lives for all available prior May 2006 data sets have been produced at the National Nuclear Data Center (NNDC) 
and publicly accessible from the NNDC website {\it http://www.nndc.bnl.gov/bbdecay} \cite{08Pr}. NNDC $^{128}$Te half-life value (3.5$\pm$2.0)x10$^{24}$ y 
is almost twice as large than that of Barabash (1.9$\pm$0.4)x10$^{24}$ y  \cite{10Ba} and NNDC  $^{128,130}$Te half-life values can be explained using 
the simple formalism of Primakoff and Rosen \cite{69Pr}. In tellurium isotopes, where both nuclei are very similar from the nuclear structure point of view, 
the difference in T$_{1/2}$ values can be attributed to differences in the $\beta\beta$-decay transition energies. 
Further analysis \cite{08Pr} indicates that 
\begin{equation}
\label{myeq.Tel0}  
T_{1/2}^{2\nu} (^{128}Te) / T_{1/2}^{2\nu} (^{130}Te) \approx 5.7 \times 10^{3} \sim (\frac{E_{130}}{E_{128}})^{8}
\end{equation}
or 
\begin{equation}
\label{myeq.Tel}  
T_{1/2}^{2\nu} (0^{+} \rightarrow 0^{+}) \sim \frac{1}{E^{8}}
\end{equation} 
This agrees well with the Primakoff and Rosen prediction of $\sim \frac{1}{E^{8.4}}$ \cite{69Pr} and provides an indication 
of the two-nucleon mechanism of  $\beta\beta$(2$\nu$)-decay.

In conclusion, the nuclear data evaluation is a fair and impartial judgment of all available data experimental results often conducted in the 
6-12 years time intervals. Evaluation policies may provide preferences to the model-independent methods over model-dependent, however, 
discrepant data sets from the same class of measurements are always included in the evaluation process using the standard statistical 
procedures such as LWEIGHT \cite{Lwei} to deduce evaluated or recommended numbers. Deviation from the nuclear structure evaluation 
policies produced underestimated T$_{1/2}$ value for $^{128}$Te \cite{00Ba} and distorted tellurium ratio for evaluated T$_{1/2}$.

\section{Acknowledgments} 
The author is grateful to B. Singh (McMaster Univversity) and M. Blennau (BNL) for  productive discussions and  help with the manuscript, respectively. 
This work was funded by the Office of Nuclear Physics, Office of Science of the U.S. Department of Energy, under Contract No. DE-AC02-98CH10886 
with Brookhaven Science Associates, LLC.  


\end{document}